\documentclass[epj]{svjour}
\usepackage{graphics}
\usepackage{amsmath}
\usepackage{xcolor}
\usepackage{amsfonts}
\usepackage{amssymb}
\usepackage{hyperref}
\usepackage{doi}
\begin{document}

\title{M2-branes, Higher Form Symmetries and 1-Gerbes}

\title{M2-branes, Higher-Form Symmetries and 1-Gerbes}

\titlerunning{M2-branes and higher-form symmetries}

\authorrunning{F. Caro-Pérez et al.}

\author{Fabián Caro-P\'erez\textsuperscript{1$\star$},
María Pilar García del Moral\textsuperscript{2$\dagger$} and
Álvaro Restuccia \textsuperscript{1$^\#$}%
\thanks{$\star$ \href{mailto:email1}{\small fabian.caro.perez@ua.cl}\,,
$\dagger$ \href{mailto:email2}{\small m-pilar.garciam@unirioja.es}
$\#$ \href{mailto:email3}
{\small
alvaro.restuccia@uantof.cl}\,}
}

\offprints{}         

\institute{{\bf 1} Departamento de Física, Universidad de Antofagasta,\\ Universidad de Antofagasta, Campus Coloso, Aptdo 02800, Antofagasta, Chile\\
{\bf 2} Area de Física, Departamento de Química,Universidad de La Rioja ,\\
e Instituto de Computación Científica de la Universidad de La Rioja (SCRIUR)\\
 C/ Madre de Dios 53, Logroño 26006, España.}

\date{Received: date / Revised version: date}

\abstract{
Higher-Form Symmetries (HFS) of a closed bosonic M2-brane formulated on a compactified target space
$\mathcal{M}_9 \times T^2$ are investigated. We show that there is an obstruction to the gauging of these global symmetries in the presence of background fields, a mixed 't~Hooft anomaly. Its cancellation is obtained by the inflow term constructed in terms of gauge fields which are flat connections on a $U(1)$-principal bundle and a torsion $\mathcal{G}_1^{\nabla_c}$-gerbe on the M2-brane worldvolume. The effect of these gauge structures together with non trivial \textit{winding} embedding maps ensures the  breaking of the continuous HFS $U(1)$ symmetry to a discrete subgroup and a worldvolume flux condition on the M2-brane.
The resulting topological operators realize discrete symmetries associated with the \textit{winding} and the flux/\textit{monopole} sectors, and their operator algebra is well-defined: the \textit{monopole} operator acts non trivially on a \textit{vortex-dressed} operator, while the winding operator acts on the transgression of the Wilson surface.
\PACS{
      {PACS-key}{M-theory, Higher Forms Symmetries, M2-brane, Gerbes, Gauge Theory}
      } 
}

\maketitle

\section{Introduction and summary}
\label{sec:intro}
In this paper, we study the existence of higher form symmetries of degree p (HFS(p)), t'Hooft anomaly cancellation, and the presence of gerbe structures on the bosonic M2-brane. The presence of these gerbe structures induces two effects. Firstly, it leads to the breaking of global symmetries into discrete gauged symmetries. Secondly, it is related to the emergence of flux quantization conditions over the bosonic M2-brane. This final result is a necessary condition for the consistency of the gauged theory in toroidal backgrounds. It is noteworthy that  the spectral analysis of the supersymmetric M2-brane compactified on toroidal backgrounds also requires the imposition of a flux condition  for the discreteness of the mass spectrum of the toroidally compactified supermembrane, 
\cite{GarciaDelMoral:2018jye}, \cite{GarciadelMoral:2020dfs}. 
\newline
Symmetries are of fundamental importance in physics. They underlie the unification of fundamental interactions, constrain the dynamics of quantum field theories, and have been used for the classification of  phases of matter. 
Conventionally, a symmetry is defined as a transformation acting on the fields of a physical system under which, the action or the equations of motion remain invariant. Such transformations are characterized by groups, finite- or infinite-dimensional. For the case of continuous symmetries Noether’s theorem guarantee the existence of conserved currents and charges generated by local operators. In a more general framework, these symmetries are often referred to as 0-form symmetries, reflecting the fact the symmetry operators act on point-like objects.
Despite their success, Noether symmetries do not characterize the full symmetry structure of quantum field theories. A limitation of Noether symmetries is that they are unable to indicate the presence of spontaneous symmetry breaking or phase transitions; for instance, transitions from deconfinement to confined phases in quantum field theory (QFT). These limitations motivated the development of a more general notion of symmetry, denoted   as \textit{Generalized Symmetries} in their seminal paper, \cite{Gaiotto:2014kfa}.
\newline
Generalized symmetries also known as Higher Form Symmetries, (HFS) represent a vast concept that generalizes the notion of symmetry \cite{Gaiotto:2014kfa}. The operator representing the HFS(p) are non-local and acts on extended charged objects (of dimension $p$, larger than zero), as for example, Wilson lines, Wilson surfaces, branes etc.  The symmetry operators are defined on submanifolds of codimension $p+1$. They extend the notion of Noether symmetry and Ward identities to extended objects instead of considering point-like sources. Furthermore, the breaking of HFS to their discrete subgroup may allow the characterizations of the phases such as confinement, Higgs or topological phases in the context of High Energy physics
but also in condensed matter physics, see for example \cite{Armas:2023tyx}. Symmetry operators are topological and gauge-invariant operators. They are classified as invertible when they satisfy group-like fusion rules.
\newline
In the context of unification theories, global symmetries are not expected to appear in any consistent quantum gravity theory \cite{Banks:2010zn}. The gauging of these HFS is realized through the introduction of gauge \textit{Background fields} which appear in the action of the theory as a  topological field term but can be dynamical in manifolds of higher dimensions.
\noindent
In the invertible cases, there is an obstruction to the gauging of the HFS, that implies the existence of anomalies in the path integral, whose cancellation involves the inclusion of new terms constructed with these background fields, like the anomaly inflow terms \cite{Kapustin:2014tfa},\cite{Cordova:2018cvg}.
\newline
To illustrate this point, consider four-dimensional Maxwell theory with its two one-form symmetries. It is a consistent theory.  However, when gauging these global symmetries a mixed ’t Hooft anomaly becomes
manifest, which can be cancelled by coupling the theory to appropriate background fields, namely higher-form gauge
connections \cite{Gaiotto:2014kfa}
In a different context, not related to a
Maxwell theory, in Supergravity the ABJ anomaly emerges due to the non-invariance of the topological operators under the global HFS symmetry, as a consequence of the \textit{Chern-Simons} term. The generators of the HFS are non-invertible \cite{Garcia-Valdecasas:2023mis}.
The original symmetry groups in the presence of sources generally break into their discrete subgroups. 
\newline
Recently, there has been a growing interest in exploring the potential applications of HFS to various gauge and gravitational theories.
See for example in the context of Supergravity {\cite{Zhang:2023wai}, string theory \cite{Bhardwaj:2023kri}, \cite{Heckman:2024obe}, F-theory \cite{Cvetic:2021sxm} and M-theory \cite{Albertini:2020mdx} among others. 
The potential applications of the study of the HFS have yet to be fully realized, with the possibility of further applications emerging.
\newline
The present paper focuses on M-theory, adopting the M2-brane as the fundamental object of study. We search for topological  effects related to HFS without introducing Supergravity dynamics. Consequently, our analysis only considers the cases of $C_{[3]}=0$ or a local 3-form $C_{[3]} $ flat connection. In any of these two cases the $C_3$ does not introduce any effect in the equations of motion of the M2-brane. The topological effects on the M2-brane are going to be captured by the gauge background fields.  
Our results indicate that the cancellation of the mixed t'~Hooft anomaly due to the gauging of the M2-brane HFS in the presence of background fields, is a direct consequence of the coexistence of two gerbe structures on the M2-brane worldvolume. In the case in which $C_{[3]}$ is a flat connection,  the background fields can be understood as induced by the pullback of a transgression form of  $C_{[3]}$ acting on the compactified sector of the target space. In the formulation discussed in section 3.2 the relevant geometrical objects characterizing the theory are, 
\begin{itemize}
  \item[i)] a on-shell flat $\mathcal{G}_1^{\nabla_c}$-gerbe \footnote{A n-gerbe is denoted $\mathcal{G}_n$-gerbe and a n-gerbe with with connection $\mathcal{G}_{1}^{\nabla}$-gerbe. The $\nabla$ represents the connection data in terms of the local forms $\nabla= \{A_{n+1}, A_n,A_{n-1}, \cdots
  \}$. For the particular case, when the connection is flat we use $\mathcal{G}_{n}^{\nabla_c}$-gerbe where c denotes close.}, classified by the torsion subgroup (and encoded by discrete \newline holonomies/Wilson surfaces);
  \item[ii)] a non-flat (twisted) $\mathcal{G}_1^{\nabla}$-gerbe whose connective data induces the worldvolume flux-quantization condition.
\end{itemize}
Together, these structures break the continuous U(1) symmetries of the M2-brane into discrete subgroups and lead to an interesting physical implication, namely a modification of the spectral properties for specific compact sectors of the target space. 
\newline
In string theory, the fundamental objects are extended objects such as strings and Dp-branes, while in M-theory they include M2-branes, M5-branes, and other M-branes.
It was known since the early nineties that bosonic $\tilde{p}$-branes with $\tilde{p}\ge 2$ are quantum inconsistent for any dimension, in clear contrast to the case of the bosonic string, which is consistent for $d=26$. The theory contains Lorentz anomalies due to order ambiguities at the quantum level, which cannot be eliminated in any gauge, not even in the Light Cone Gauge (LCG) \cite{Bars:1987nr}, \cite{Bars:1989ba}. This problem is avoided by the introduction of supersymmetry. The M2-brane also called supermembrane was discovered in the early eighties by \cite{Bergshoeff:1987cm}. The supersymmetric M2-brane can be formulated consistently on certain dimensions, \textit{i.e.} in $d=4,5,7$ and $11$ dimensions as demonstrated in \cite{Duff:1987cs}. At the supersymmetric level, the action can be made Lorentz anomaly free for $D=11$, by considering its formulation on a Light Cone Gauge (LCG) and imposing kappa symmetry, \cite{Marquard:1989rd}. The resulting theory, formulated \textit{á la} Green-Schwarz, is invariant under area preserving diffeomorphism. This is the  residual symmetry of the M2-brane in the LCG formulation. 

\noindent
At quantum level the supermembrane behaves very differently to the string. The first quantization of a String theory is formulated in terms of the one-string Hamiltonian. The center of mass of a string moving in Minkowski spacetime is continuous, but the internal structure of the theory is described by the internal spectrum of the one-string Hamiltonian. The discreteness of its spectrum is what gives the theory its interpretation as a tower of mass excitations. If the spectrum were continuous, it would be harder to interpret the theory as a clean sequence of quantum one-object particle excitations.

\noindent
The situation is completely different in the case of supermembrane theory. Classically, when formulated on a Minkowski space-time, the M2-brane bosonic potential contains flat directions that admit string-like spikes as non-trivial solutions with zero energy cost. These spikes are responsible for the non-conservation of the M2-brane's topology and the non conservation of the number of particles. 
In $1988$, the matrix model of the supermembrane formulated on Minkowski space-time was discovered, \cite{deWit:1988wri}. The supersymmetric mass spectrum of this model was rigorously proven to be continuous from zero to infinity \cite{deWit:1988xki}. This result changed the direction of the field, with matrix theory being reinterpreted as a second quantised theory defined in terms of multiple D0-branes, \cite{Banks:1996vh}. This corresponds to the regularised supermembrane and to the $0+1$-dimensional reduction of 10D Super Yang-Mills theory.
These important results, together with the fact that toroidal compactification did not alter them, led the scientific community to believe that, unlike the string case, supermembrane theory did not admit a first quantised description.

\noindent
However, although this is true in many cases, four different backgrounds have been obtained on which the supermembrane theory has a purely discrete spectrum with finite eigenvalues. These correspond in first place, to a supermembrane  toroidally compactified with a central charge induced by an irreducible wrapping (see \cite{Boulton:2002br},\cite{Boulton:2010nd}). The second example corresponds to a supermembrane formulated on a pp-wave \cite{Dasgupta:2002hx}. The latter's matrix model corresponds to the well-known BMN matrix model (see \cite{Berenstein:2002jq}). In \cite{Boulton:2010nd} it was proven that the matrix model spectrum is discrete beyond the semiclassical limit. 
The third case corresponds to a supermembrane formulated on a toroidally compactified space, $\mathcal{M}_9\times T^2$, in the presence of quantized flat Supergravity 3-form $C_{[3]} $, see \cite{GarciaDelMoral:2018jye} \cite{GarciadelMoral:2020dfs}. The last one discovered corresponds to an M2-brane on a twice-punctured Riemann surface times Minkowski space, describing a massive supermembrane, \cite{delMoral:2021fvw}.
\newline
It is reasonable to conjecture that these behaviors may emerge as a consequence of the cancellation of a non previously considered anomaly of the M2-brane.  This possibility has not been considered in the past. Our analysis indicates that this is indeed the case: The partition function of the membrane, once the action has been gauged by the background fields, exhibits a 't Hooft anomaly. This anomaly, signaling an obstruction to gauging the theory, is canceled by introducing an inflow term defined in terms of background fields supported on a four-dimensional manifold, which couple to the three-dimensional membrane action through a BF coupling.
\newline
This anomaly cancellation gives rise to geometrical structures of a higher principal $U(1)$-bundle also called  $\mathcal{G}_1^{\nabla}$-gerbes on the M2-brane theory. The presence of these gerbes have two important consequences, on one hand the breaking to a discrete subgroup of the global symmetry and in the second place generating non-trivial fluxes over the M2-branes. The projection of this structure through the worldvolume gives rise to a flux term on the membrane. The flux quantization condition guarantees the quantization of the charges and is associated with the existence of a \textit{monopole} over the membrane. It is noteworthy that the supermembrane with worldvolume fluxes, i.e. \textit{monopoles} compactified on a toroidal background \cite{Martin:1997cb}, has a purely discrete supersymmetric spectrum with finite multiplicity \cite{Boulton:2002br}, \cite{Boulton:2010nd}. This property also holds in the presence of quantized three-forms whose pullback induces, the worldvolume flux conditions over the M2-brane \cite{GarciaDelMoral:2018jye}, \cite{GarciadelMoral:2020dfs}.
\newline
The paper is structured as follows: in Section \ref{Capitulo HFS} we briefly review HFS. In Section \ref{CHAPTER: HIGHER FORMS SYMMETRIES MEMBRANE} we present new results: we obtain the HFS of the bosonic M2-brane theory which cancels the 't Hooft anomaly in 11D. {The breaking of the U(1) symmetries into its discrete subgroups is shown explicitly.  In \ref{CHAPTER: TORSIONAL SECTOR OF M2-BRANE} the
explicit construction of the torsional sector of the bosonic M2-brane and its associated gerbe structures are presented. The coupling of the winding current to this sector is demonstrated, explicitly showing the appearance of a non-trivial Wilson surface on the M2-brane, which corresponds to the holonomy of the flat $\mathcal{G}_1^{\nabla_c}$-gerbe. In Section \ref{CHAPTER: ALGEBRA} the algebra of operators is found. The action of the monopole operator over a dressed vortex operator is obtained as well as the action of the winding operator over a pullback of the charged Wilson Surface. In Section \ref{CHAPTER: CONCLUSIONS} we present a  brief discussion and our conclusions. In appendix \ref{APENDICE DEMOSTRACION FREE ANOMLA TERM}, the anomaly cancellation of the M2-brane path integral in the presence of background fields is shown.

\section{Higher form symmetries for gauge theories}\label{Capitulo HFS}

In this section, we briefly introduce HFS(p).
In a standard QFT, the Noether symmetries of the theory are associated with unitary operators of codimension-1 acting on charged point-like objects. To generalize this, one considers topological operators of codimension-$(p+1)$ associated to the symmetry group $\mathbb{G}^{(p)}$ acting on p-dimensional charged objects. We denote them by $U_g$ where the label $g \in \mathbb{G}^{(p)}$ \cite{Bhardwaj:2023kri}, \cite{Gaiotto:2014kfa}, \cite{Gomes:2023ahz}, \cite{Schafer-Nameki:2023jdn}. These operators are topological if they remain invariant under smooth deformations of the integration submanifold, as follows:
\begin{align}\label{OPERADORES GENERALIZADOS}
    U_{g}(\Sigma_{d-p-1}) = U_{g}(\eta(\Sigma_{d-p-1})),
\end{align}
Where $\eta: \Sigma_{d-p-1}\to \overline{\Sigma}_{d-p-1}'$, where $\overline{\Sigma}'_{d-p-1}$ has the same orientation of the original manifold.
This is only possible if there exists a $(d-p)$- manifold $\mathcal{Y}_{d-p} \in \Omega_{0}$
\begin{align}
    \partial \mathcal{Y}_{d-p} = \Sigma_{d-p-1} \sqcup \Sigma_{d-p-1}',
\end{align}
where \( \Sigma_{d-p-1}' \) has the opposite orientation with respect to $\Sigma_{d-p-1}$.
Furthermore, the operators
in \eqref{OPERADORES GENERALIZADOS} act on each other according to fusion rules inherited from the \textit{symmetry} group of the field
theory formulated on the base manifold. 
Let $g, g'\hspace{1mm} \& \hspace{1mm} gg' \in G$, then we have:
\begin{align}\label{PROPIEDAD INVERTIBILIDAD}
    U_{g}(\Sigma_{d-p-1}) U_{g'}(\Sigma_{d-p-1}) = U_{gg'}(\Sigma_{d-p-1}), 
\end{align}
where $U_g \in \mathbb{G}^{(p)}$. $\mathbb{G}^{(p)}$ is the global symmetry group associated to the HFS(p), being $p$ the degree of the HFS where the topological operators act.
\newline
The symmetry operators act by automorphisms on the algebra of the charged operators $\mathcal{O}$ as follows: 
\begin{align}\label{objetocargado}
    U_{g}(\Sigma_{d-p-1}) ~\mathcal{O}~ U_{g}(\Sigma_{d-p-1})^{-1} = g\cdot\mathcal{O}, 
\end{align}
In a QFT, we have two possible types of HFS: invertible and non-invertible symmetries. A generalized symmetry is invertible if the previous topological operators satisfy \eqref{PROPIEDAD INVERTIBILIDAD}, \eqref{objetocargado}. 
A non-invertible symmetry is generated by a topological operator that does not admit an inverse under fusion; its fusion rules are not of group-like (\ref{PROPIEDAD INVERTIBILIDAD}). Fusion rules typically form a category and its action on the operator algebra fails to be an automorphism. For a discussion see, \cite{Garcia-Valdecasas:2023mis}, \cite{Karasik:2022kkq}, \cite{Choi:2022jqy}, \cite{Fernandez-Melgarejo:2024ffg}, \cite{Hasan:2024aow}, \cite{Hull:2024bcl}, \cite{Cvetic:2023plv} and \cite{Luo:2023ive}. 
\newline
In this paper we are only concerned with invertible symmetries.
\newline
The invertible HFS$(p)$ are obtained by calculating the generalization of the Noether current 
$\star J_{d-p-1} \in \Omega^{p+1}(\mathcal{M})$, whose associated form is closed in the \textit{de Rham} cohomology sense. This is equivalent to requiring that the current follows a conservative law of structure \textit{á la} Noether:
\begin{align*}
   \star J_{d-p-1} \in Z^{p+1}(\mathcal{M},\mathbb{R}).
\end{align*}
Using the map from the Lie groups to its associated Lie algebras: $ G\to Alg(G)=\mathfrak{g}$ with $g=e^{i\alpha}$, we can define the invertible topological operator as follows \cite{Gaiotto:2014kfa}:
\begin{align}\label{OPERADORES U}
      U_{\alpha}(\Sigma_{p+1}) \triangleq {}& \exp \big(i\alpha \int_{\Sigma_{p+1}} \star J_{d-p-1}  \big).
\end{align}
In four-dimensional Maxwell theory \cite{Gaiotto:2014kfa}, \cite{Gomes:2023ahz}, there are an electric and magnetic 1-form symmetries, i.e.\ HFS$(1)$. The gauging the global HFS$(1)$ symmetry makes manifest a mixed ’t~Hooft anomaly, in fact, the transformation of the path integral $\mathcal{Z}(A,\{B_I\}_I)$ under the 1-form gauge transformation is
\begin{align}
    \mathcal{Z}([A], \{[ B _{I} ]\}) \;=\; \phi(\Lambda, \{ B_{I}\}) \, \mathcal{Z}(A, \{B_{I}\}) ,
\end{align}
where $\phi(\Lambda,\{B_I\})$ is a $U(1)$ phase. This phase encodes the mixed anomaly between the electric and magnetic 1-form symmetries. Here $\{B_I\}$ denotes the collection of background gauge fields associated with the $I$-th HFS$(p)$ symmetry of the theory \cite{Schafer-Nameki:2023jdn}. The anomaly reflects the fact that one cannot gauge the electric and magnetic 1-form symmetries simultaneously.
To restore gauge invariance, one must introduce an anomaly-inflow action, governed by an invertible anomaly theory (a TQFT) defined on a 5d-manifold $\mathcal{D}$ up to bordisms, with boundary
$\partial\mathcal{D}=\mathcal{M}$. This motivates defining a modified path integral $\widetilde{\mathcal{Z}}$, in which the TQFT factor cancels out the anomalous contribution.

\section{Higher form symmetries  of the bosonic M2-brane }\label{CHAPTER: HIGHER FORMS SYMMETRIES MEMBRANE}

The action of a bosonic $\tilde{p}$-brane\footnote{There is no correlation between the   $\tilde{p}$ and the $p$ appearing in HFS(p).}, \textit{á la} Nambu Goto \cite{Goto:1971ce}, formulated on a flat spacetime $\mathcal{M}_{11}$ is, given by: 
\begin{equation}
S_{NG}(X)=-\frac{T_{M\tilde{p}}}{2}\int_{\Sigma_{\tilde{p}+1}} d^{{\tilde{p}}+1}\xi[ \det(\partial_i X^{M}\partial_jX_{M})]^{\frac{1}{2}},
\end{equation}
where the dynamical fields correspond to the embedding maps from the $\tilde{p}$-brane worldvolume $\Sigma_{\tilde{p}+1}$ into the target space: $$X^{M}:\Sigma_{\tilde{p}+1}\to \mathcal{M}_{11},$$ where $\Sigma_{\tilde{p}+1}$ is a closed Riemaniann manifold of $\tilde{p}+1$ dimensions and $\mathcal{M}_{11}$ is a space-time manifold with a flat metric which generically will contain compact sectors and nontrivial cycles. Specifically, we will consider  $\mathcal{M}_{11}=
\mathcal{M}_{11-q}\times T^q$ being $\mathcal{M}_{11-q}$ a euclidean version of Minkowski spacetime of dimension $11-q$ and $T^q$ representing a q-torus. The index $M, N,...=\{ 0, \cdots, 10\}$.
We denote by $\xi^i$, the local coordinates on $\Sigma_{\tilde{p}+1}$. $T_{M\tilde{p}}$ is the tension of the $\tilde{p}$-brane. The maps  $X^M(\xi^i)$ transform as a vector on the target space and as scalars on the worldvolume. The bosonic M2-brane action is obtained by specializing to the case $\tilde{p}=2$, being $\Sigma_3$, the $(2+1)$  M2-brane worldvolume. The M2-brane acts as a source for 11D Supergravity. It also admits a Polyakov type action, see \cite{Polyakov:1981rd} for further details. In addition to the embedding maps, an independent auxiliary metric, $g_{ij}$, with determinant $g$, is included in the worldvolume action,
\begin{equation}\label{ACTION MEMBRANE}
    S_P(g,X)=-\frac{1}{2}T_{M\tilde{p}}\int_{\Sigma_{\tilde{p}+1}}d^{\tilde{p}+1}\xi \bigg( \sqrt{g}g^{ij}\partial_iX^{M}\partial_jX_{M}-\sqrt{g}\Lambda   \bigg).
\end{equation}
The equation of motion obtained from the variation of the auxiliary metric yields an induced metric in terms of the embedding maps.
\begin{equation}\label{induced}
    g_{ij}=\frac{(\tilde{p}-1)}{\Lambda}\partial_iX^{M}\partial_jX_{M}.
\end{equation}
In the following we take $\Lambda=\tilde{p}-1$, hence, $\Lambda=1$ in the case of the membrane $\tilde{p}=2$. This action can also be coupled to a \textit{Wess-Zumino} term via the coupling to the pullback $\mathbb{P}$ of the Supergravity three-form $C_{[3]}$
\begin{align}
    \frac{T_{M2}}{2}\int_{\Sigma_3}\mathbb{P}(C_{[3]})= 
    \frac{T_{M2}}{2}\int_{\Sigma_3} \widehat{C}_{ijk} d\xi^i\wedge d\xi^j\wedge d\xi^k,
\end{align}
where $ \hat{C}_{ijk}$ is the coefficient of the pullback of the $11D$ bosonic part of the Supergravity three-form $C_{[3]}$. In this paper, we study the effect of the cancellation of the 't Hooft anomaly at the level of the bosonic membrane without the WZ term.
This term will be considered in a subsequent work.  Our results are also valid when the WZ term is expressed in terms of a flat quantized 3-form $C_3$ connection. In this case the topological properties of $C_3$ have to be compatible with the Gerbe structure of the background fields.  Therefore, we will not consider here the introduction of this important term any further. 
\newline
The symmetries of this action are, a local symmetry associated with the invariance under general coordinate transformations \cite{Bergshoeff:1987cm}, and the invariance under a global translational symmetry \footnote{In  \cite{Chatzistavrakidis:2021dqg} in the context of String theory a similar transformation was considered to generate a HFS.}, 
\begin{equation}\label{SYMMETRIA GLOBAL MEMBRANA ANAISADA}
X^{M}\to X^{M}+\epsilon^{M}.
\end{equation}
This is the symmetry that we will be interested in this study.  We will distinguish the embedding maps $X^r$ to the compact sector of the target space from the maps to the non-compact sector of the target, $X^m$. The translation on the coordinates of the latter implies the conservation of the linear momentum of the center of mass of the membrane, while the shift on the $X^r$ coordinates can be interpreted $U(1)$ symmetries.
By considering the Polyakov type action, we can formulate it in the language of forms as follows 
\begin{equation}\label{accion s}
    S_{P}(g,X)=-\frac{T_{M_2}}{2}\int_{\Sigma_3} dX^{M}\wedge \star dX_{M}-\frac{T_{M_2}}{2}\int_{\Sigma_3}\sqrt{g} \omega_{vol}(\Sigma_3) ,
\end{equation}
where the Hodge dual is defined with respect to the worldvolume $\Sigma_3$ metric,
\begin{align*}
    \star dX=\sqrt{g}\epsilon_{jkl}g^{ji}\partial_iXd\xi^k\wedge d\xi^l 
\end{align*}
and $\omega_{vol}(\Sigma_3)$
is a volume-form in the worldvolume. In the symmetry analysis, we will consider $\Sigma_3$ to be a compact manifold without boundary (a closed manifold).
The second term in \eqref{accion s} will be denoted $\tilde{S}(g)$.
It is relevant since it is involved in the determination of the metric $g_{ij}$.
\newline
The equation of motion arising from the variation of the action \eqref{accion s} with respect to $X^{M}$, together with the \textit{Bianchi} identity are:
\begin{align}\label{EQ DE MOVIMIENTO POLYAKOV}
    d\star dX^{M}={}&0, \\
    d(d X^{M})={}&0 \label{EQ DE BIANCHI MEMBRANA},
\end{align}
where the Hodge dual is evaluated with respect to the induced metric given in \eqref{induced}, with $\Lambda=1$. 
Following the standard analysis for HFS(p), we reinterpret \eqref{EQ DE MOVIMIENTO POLYAKOV} and \eqref{EQ DE BIANCHI MEMBRANA} as the conservation law of higher currents. For this case, we have the following set of currents, symmetries,
\begin{align}\label{CORRIENTES DE HFS(0) PARA POLYAKOV}
    J_{m}^{M} = \star dX^{M}, \qquad J_{w}^{M} = dX^{M}.
\end{align}
The sub-index $m$ and $ w$, in the case when the target space is compact and the associated the currents are $J_m^r$, $J_w^r$ they denote \textit{monopole} and \textit{winding} currents, respectively. The precise meaning will become evident soon. They are associated with the HFS(0) and HFS(1), respectively.
These $p-$form currents are conserved by construction. This global symmetry can be seeing as generated by the topological operator built as:
\begin{align}\label{OPERADORES DE M2 BRANA WINDING SIN B}
     U_{\beta}^{r}(\mathcal{Q})={}&\exp\big(i\beta \int_{\mathcal{Q}}dX^{r} \big),
\end{align}
where $\mathcal{Q}$ is a closed 1-submanifold of $\Sigma_3$, i.e. compact without boundary. One can observe that we have restricted to the compact sector since it is the only one that generates the non-trivial $U(1)$ symmetries.
When there are non-trivial cycles in the compact sector of the target space, the embedding maps onto the compact sector can be interpreted as angular variables.  Since the shift defines an abelian group isomorphic to a $U(1)$ group whose parameter is $\beta$ and it is associated to the \textit{winding} of the membrane on the compact sector of the target space. The other topological operator associated to \eqref{EQ DE MOVIMIENTO POLYAKOV} is the following one
\begin{align}\label{OPERADORES DE M2 BRANA SIN B}
    U_{\alpha}^{r}(\mathcal{N})={}&\exp\big( i\alpha\int_{\mathcal{N}} \star dX^{r}\big),
\end{align}
where $\mathcal{N}$ is a closed 2 submanifold of $\Sigma_3$. From \eqref{EQ DE MOVIMIENTO POLYAKOV} we conclude that $\star dX^r$ is a closed two-form which, when its periods are integers, they can be interpreted as the curvature of a one-form connection on $\Sigma_3$ associated to a principal $U(1)$-bundle. We call it briefly a \textit{monopole} $U(1)$ bundle. $\alpha$ is the parameter associated to this $U(1)$ symmetry. 
\newline
The operators defined in \eqref{OPERADORES DE M2 BRANA WINDING SIN B} and \eqref{OPERADORES DE M2 BRANA SIN B} naturally act on the states of the theory. These operators are topological.
The equations of motion \eqref{EQ DE MOVIMIENTO POLYAKOV} and \eqref{EQ DE BIANCHI MEMBRANA} hold on the worldvolume of the membrane $\Sigma_3$, hence they are also valid on
the submanifolds $\mathcal{N}$ and $\mathcal{Q}$.
The action of these operators is trivial unless the submanifolds  $\mathcal{Q}$ and $\mathcal{N}$ are non-trivial cycles, and the associated 1- and 2-forms are harmonic forms.
\newline
The global symmetries of M2-brane are
\begin{align}
\mathbb{G}^{(1)}\cong\bigoplus_{i=1}^qU(1)^{(1)}_{w,i}  \oplus U(1)_{m,i}^{(0)}.
\end{align}
Where each sector corresponds to the \textit{winding} and \textit{monopole} contribution, respectively. They are associated to dual currents in 3D in the sense of \cite{Martin:1994np}.
\subsection{Background field coupling and the free-anomaly action}\label{SECCION: Background field coupling and the free-anomaly action}

In this and the following sections we are going to analyze the gauging of the global symmetries. In order to do it we introduce auxiliary fields coupled with the currents \eqref{CORRIENTES DE HFS(0) PARA POLYAKOV} in the action. In this section they are gauge fields with integer periods, $(\mathcal{B}^r,\Tilde{\mathcal{B}}^r)\in \Omega^1(\Sigma_3)\oplus\Omega^2(\Sigma_3)$ are U(1) gauge fields.
On the other hand, 
In the following, we will consider the worldvolume $\Sigma_3=T^2\times S^1$ with the embedding maps coupled to the currents (\ref{CORRIENTES DE HFS(0) PARA POLYAKOV}). We start considering,
\begin{align}\label{ECAUCION: ACCION LIBRE DE ANOMALIA}
     S_{}(g,X^r,X^m, \mathcal{B}^r,\Tilde{ \mathcal{B}}^r)={}&-\frac{T_{M2}}{2}\int_{\Sigma_3}(dX-\mathcal{B})^r\wedge \star (dX-\mathcal{B})_r +\notag
     \\
     {}&+  \frac{T_{M2}}{2}\int_{\Sigma_3}\Tilde{\mathcal{B}}^r\wedge(dX-\mathcal{B})_r+\notag
     \\
     {}&
     +S_{nc}(g,X^m)+\widetilde{S}(g).
\end{align}
Where $\widetilde{S}(g)=-\frac{T_{M2}}{2}\int_{\Sigma_3}\sqrt{g}\omega_{vol}(\Sigma_3)$, is the worldvolume term in \eqref{ACTION MEMBRANE} and 
\begin{align*}
    S_{nc}(g,X^m)= -\frac{T_{M2}}{2}\int_{\Sigma_3}dX^m\wedge \star dX_m
\end{align*}
is associated to the non-compact embedding maps. In order to gauge the global symmetry, the 1-form class $ [\mathcal{B}^r]$ must transform in terms of an exact form defined with the same 0-form parameter as in the \eqref{SYMMETRIA GLOBAL MEMBRANA ANAISADA}. The gauge transformations are,\
\begin{align}
   [X^r] {}& \triangleq  X^{r} \to X^{r}+\epsilon^r,  \label{CLASE DE EQUIVALENCIA X B0}\\
   [\mathcal{B}^r] {}&  \triangleq \mathcal{B}^{r} \to
     \mathcal{B}^{r}+d\epsilon^{r}, \label{CLASE DE EQUIVALENCIA X B1}\\
    [\tilde{\mathcal{B}}^r] {}& \triangleq \Tilde{\mathcal{B}}^{r} \to \Tilde{\mathcal{B}}^{r}+d\Lambda_1^{r}.\label{CLASE DE EQUIVALENCIA X B2}
\end{align}
The introduction of these background fields has provided compelling evidence for the 't Hooft anomaly given by:
\begin{align}
    \delta S_{}(g,X^r,\mathcal{B}^r,\Tilde{ \mathcal{B}}^r)=-\frac{T_{M2}}{2}\int_{\Sigma_3}\mathcal{B}^r\wedge d\Lambda_{1r}.
\end{align}
If we introduce a counterterm for this anomaly, as follows
\begin{align}
    S_{ct}(\mathcal{B}^r,\tilde{\mathcal{B}}^r)={}& \frac{T_{M2}}{2}\int_{\Sigma_3}\mathcal{B}^r\wedge\tilde{\mathcal{B}_r},
\end{align}
the variation of the action containing  the former counterterm, still contains an anomaly associated with the variation:
\begin{align}
    \delta \big( S_{}(g,X^r,\mathcal{B}^r,\Tilde{ \mathcal{B}}^r)+S_{ct}(\mathcal{B}^r,\Tilde{ \mathcal{B}}^r) \big) ={}&\frac{1}{2}\int_{\Sigma_3}\Tilde{\mathcal{B}}^r\wedge d\epsilon_r.
\end{align}
It is evident here that the problem of the 't Hooft anomaly arises because the introduction of counterterms exchanges the anomalous term between \((w) \to (m)\) and \((m) \to (w)\). This is similar to how, in \(U(1)\) gauge theory, counterterms interchange the roles of electric and magnetic anomalies \cite{Brennan:2022tyl}. In the present case, the required anomalous counterterm is given by an inflow action with a BF structure \cite{Restuccia:1998yx} defined in a four dimensional manifold $\mathcal{D} \in \textbf{Bord}_4$ such that $\partial \mathcal{D}=\Sigma_3$:
\begin{align}
   \mathcal{T}_{TQFT}(\mathcal{B}^r,\Tilde{\mathcal{B}}^r) ={}& \frac{T_{M2}}{2}\int_{\mathcal{D}}\mathcal{B}^r\wedge d\tilde{\mathcal{B}_r}.
\end{align}
In what follows we fix $T_{M2}=1$. For every closed, orientable 3-dimensional smooth manifold $\Sigma_3$ there always exist an orientable smooth $\mathcal{D}$ such that 
\begin{align}
  \partial \mathcal{D}=\Sigma_3
\end{align}
in fact, generically $\mathcal{D}$ exists if and only if the class of the worldvolume
\begin{align}
[\Sigma_3]=0 \quad \textrm{in} \quad \Omega_3^{so},
\end{align}
The result follows since $\Omega_3^{so}=0.$ In this construction $\mathcal{D}$ is an auxiliary manifold, not a submanifold of the target. The action is independent of the choice of $\mathcal{D}$,
since $\mathcal{B}^r$ and $\widetilde{\mathcal{B}}^r$ are gauge fields, which we assume that  $d\mathcal{B}^r\in H^2(\mathcal{D},\mathbb{Z})$ and $d\Tilde{\mathcal{B}^r}\in H^3(\mathcal{D},\mathbb{Z})$. See Appendix \ref{APENDICE: INDEPENDENCIA DEL BORDISMO}. 

\noindent
In the next section we show that, although the target is a direct product, the non-trivial   topology of the background gauge fields give rise to the breaking of the $U(1)$ symmetry. 
\newline
Therefore, the anomaly-free action including the noncompact sectors is given by,
\begin{align}
    S_{T}(g,X^m, X^r,{}&\mathcal{B}^r,\Tilde{\mathcal{B}}^r)=-\frac{1}{2}\int_{\Sigma_3}DX^r\wedge \star DX_r+\notag
    \\
    {}&+\frac{1}{2}\int_{\Sigma_3} \Tilde{\mathcal{B}}^r \wedge dX_r+ \frac{1}{2}\int_{\mathcal{D}}\mathcal{B}^r\wedge d\Tilde{\mathcal{B}} ^r+ \\ \notag
    {}& +S(g,X^m)+\tilde{S}(g),
\end{align}
where \( DX^r \) is defined as \( DX^r \triangleq dX^r - \mathcal{B}^r \).  
It is straightforward to prove that the action is invariant under the equivalence class \eqref{CLASE DE EQUIVALENCIA X B0} and \eqref{CLASE DE EQUIVALENCIA X B1}. In the appendix \ref{APENDICE DEMOSTRACION FREE ANOMLA TERM}, it is explicitly shown that the inflow term directly cancels the anomaly. In the appendix \ref{APENDICE: INDEPENDENCIA DEL BORDISMO} we show that the inflow term is independent of the choice of the auxiliary four-dimensional manifold $\mathcal{D}$.

\noindent
The equations of motion of the modified action in $\Sigma_3$, from variation with respect to $X^r$ are:
\begin{align}\label{EQ DE BACKGGROUMD}
    d\star (dX-\mathcal{B})^r=0.
\end{align}and from the Bianchi identity, $  d(dX-\mathcal{B})^r=0,$ since $dB^r=0$ on-shell.
In this case the currents become
\begin{align}\label{CORRIENTES DE HFS(1)}
    J_{m}^r = \star (dX-\mathcal{B})^r, 
    \qquad J_{w}^r = dX^r-\mathcal{B}^r
\end{align}
At this stage, the symmetry group has not changed, but it has become gauged with an M2-brane action which is anomaly-free.  The background fields are dynamical gauge fields which appear in the measure of the partition function, see the appendix \ref{APENDICE DEMOSTRACION FREE ANOMLA TERM}.

\subsection{\texorpdfstring{ The M2-brane on $\mathcal{M}_9\times T^2$ with a Gerbe structure.}{The M2-brane on $\mathcal{M}_9 \times T^2$ with a Gerbe structure.}}\label{SECCION: M9T2 GERBE}
For the present analysis  we consider a particular background, $\mathcal{M}_9\times T^2$ where $\mathcal{M}_9$ is the euclidean version of the nine dimensional Minkowski space-time and $T^2$ a flat 2-torus. This background  has been used extensively in the past to analyze properties of the M2-brane. It has the virtue that is relatively simple but contains a nontrivial topology that allows the presence of \textit{monopoles}.
We consider the embedding maps from the worldvolume of the M2-brane $\Sigma_3$ into the target space,  
\begin{align}
   \Sigma_3 \longrightarrow \mathcal{M}_9 \times T^2.
\end{align}
The worldvolume of the M2-brane, $\Sigma_3$, is locally $T^2\times S^1$. Globally, however, it may be a $T^3$  or a nilmanifold forming a 3-twisted torus. In the last case there exists a non-trivial monodromy around the cycles of the worldvolume, which induce a non-trivial topology on the background gauge fields. 
The gauge fields are described in terms of flat gerbes, defining torsion classes. We will consider this case in section \ref{CHAPTER: TORSIONAL SECTOR OF M2-BRANE}. In all cases the target space remains $\mathcal{M}_9\times T^2$.

\noindent
In the section,\ref{SECCION: Background field coupling and the free-anomaly action} we have considered  $\Sigma_3$ with the $T^3$ topology on the worldvolume and background fields to be global forms on $\Sigma_3$, \textit{i.e.} in that case there is no torsion. 

\noindent
In this section, we consider the $T^3$ topology on the worldvolume and the embedding maps wrap on the cycles of the target. The background fields are locally forms but globally they behave as  $U(1)$ connections with nontrivial transitions. \textit{On-Shell}, they will end up in a torsion class.

\noindent
Now, preserving the cancellation of the ’t Hooft anomaly, we consider $\widetilde{\mathcal{B}}^r$ the connection 2-form of a $\mathcal{G}_1^{\nabla}$-gerbe. The $\mathcal{G}_1^\nabla$-gerbe are classified by the elements of $H^3(\Sigma_3,\mathbb{Z})$ \cite{Johnson:2002tc}, \cite{Murray:1999ew} \footnote{A principal \(U(1)\)-bundle equipped with (a set) connection $\nabla$ can be interpreted as the \(\mathcal G_0^\nabla\)-gerbe case in the hierarchy of \(\mathcal G_n^\nabla\)-gerbes with connection. Accordingly, \(\mathcal G_n^\nabla-\text{gerbe}\) denotes an \(n\)-gerbe with connection over $\mathcal{M}$, whose \textit{Dixmier-Douady} class lies in \(H_D^{n+2}(\mathcal{M},\mathbb Z)\). In particular,
\[
\mathcal G_0^\nabla-\text{gerbe}\longleftrightarrow H_D^2(\mathcal{M},\mathbb Z),\qquad
\mathcal G_1^\nabla-\text{gerbe}\longleftrightarrow H_D^3(\mathcal{M},\mathbb Z),
\]
so that ordinary principal bundles appear as the first level in this hierarchy \cite{Murray:1994db}.}. In the case of a  flat $\mathcal{G}_1^{\nabla_c}$-gerbe, the \textit{Dixmier-Douady} class, briefly $DD$-class, is in $Tor (H^3(\Sigma_3,\mathbb{Z}))$. $\widetilde{\mathcal{B}}$ has non trivial transitions: It is locally a 2-form on $\Sigma_3$ but not globally. The transformation $\widetilde{\mathcal{B}}\to\widetilde{\mathcal{B}}+d\Lambda_1$ in \eqref{CLASE DE EQUIVALENCIA X B2} is in agreement with the assumption that $\widetilde{\mathcal{B}}$ is a gauge connection.\newline
Also, it is natural to couple $dX^r$ with a $U(1)$-connection, as we have in the action. The $X^r, r=1,2$ are \textit{embedding} maps from $\Sigma_3\to T^2$ in the compact sector of the target space. The 1-forms $dX^r$ have then integral periods. We thus consider $[dX^r]$, a class in $H^1(\Sigma_3,k\mathbb{Z})$, where the elements of the class are related by the transformation $dX^r\to dX^r +kd\epsilon^r$. 

\noindent
The background fields  $\mathcal{B}^r\in H^1(\Sigma_3,U(1))$ are gauge connections on the $U(1)$ principal bundles, their curvatures $H^r=d\mathcal{B}^r$ belong to a Chern class $[H^r/2\pi]\in H^2(\Sigma_3,\mathbb{Z})$. \textit{On-Shell}, we will show that the $U(1)$ gauge group must breakdown to a subgroup $\mathbb{Z}_k$. 

\noindent
The total action $S_T$, once we have considered the coupling to the TQFT with level $k$, becomes:
\begin{align}\label{ECUACION: ACTION FREE ANOMALY}
S_T(g,X^r,X^m,\mathcal{B}^r, {}&\Tilde{\mathcal{B}}^r)=-\frac{1}{2}\int_{\Sigma_3}\bigg(DX^r\wedge \star DX_r-\notag
\\
{}&-dX^r\wedge \Tilde{\mathcal{B}}_r\bigg)+\frac{k}{2}\int_{\mathcal{D}}\mathcal{B}^r\wedge d \tilde{ \mathcal{B}}^r +\notag
\\
{}&+S_{nc}(g,X^m)+\tilde{S}(g).
\end{align}
Where $\mathcal{D}$ is a four dimensional auxiliary manifold with boundary $\Sigma_3$, and $DX^r=dX^r-k\mathcal{B}^r.$
The $dX^r\wedge \Tilde{\mathcal{B}}_r$ term in the integration on $\Sigma_3$ can be reinterpreted 
in terms of a bundle $\mathcal{G}_1^{\nabla}$-gerbe with a non trivial curvature when $dX^r$ wraps on the target torus,   \cite{Murray:1994db}, \cite{Murray:1999ew}. We may then lift the $\mathcal{G}_1^{\nabla_c}$-gerbe with connection $\Tilde{\mathcal{B}}^r$ to a $\mathcal{G}_1^\nabla$-gerbe with nontrivial class $[dX^r\wedge \widetilde{\mathcal{B}}^r]$.\\

\noindent
We now consider the dynamical equations associated to the previous action. We then take variations with respect to the independent fields. 
\begin{itemize}
    \item[$\bullet)$] Variations with respect to the \textit{embedding} maps $X^r$:
    \begin{align}
    d\star (dX^r-k\mathcal{B}^r)+d\tilde{\mathcal{B}}^r=0, ~~ in ~ \Sigma_3.
\end{align}
\end{itemize}
\begin{itemize}
    \item[$\bullet)$] Now, taking the variations with respect to ${\mathcal{B}}^r$:
\begin{align}\label{eq:35}
    d\tilde{\mathcal{B}^r}={}&0, ~~ in ~ \mathcal{D}.
\end{align}
\end{itemize}
Hence, if $i:\partial \mathcal{D}=\Sigma_3\longrightarrow \mathcal{D}$ is the inclusion map, then  $i^*d\tilde{\mathcal{B}^r}=0.$ That is,
\begin{align}
    d\tilde{\mathcal{B}^r}=0 ~~ in ~ \Sigma_3,
\end{align}
where $d$ denotes the exterior derivative in $\Sigma_3$ and hence,
\begin{align}
    \star (dX^r-k\mathcal{B}^r)=0 ~~ in ~\Sigma_3,
\end{align}
that is 
\begin{align}\label{eq38}
    dX^r-k\mathcal{B}^r=0 ~~ in ~\Sigma_3.
\end{align}
\begin{itemize}
    \item[$\bullet)$] Variations with respect to $\tilde{\mathcal{B}^r}$:
\begin{align}\label{eq:39}
    d{\mathcal{B}}^r=0 ~~ in~\mathcal{D},\textrm{then} \quad d{\mathcal{B}}^r=0 ~~ in ~\Sigma_3.
\end{align} 
\end{itemize}
and the same expression of \eqref{eq38} is also obtained. \\

\noindent
Consequently, \textit{On-Shell},   the \eqref{eq:35} $\mathcal{G}_1^\nabla$-gerbe becomes flat and $dX^r-k\mathcal{B}^r=0$. 
That is, \textit{on-shell} the bulk term does not see that gerbe, but the term $\int_{\Sigma_3}dX^r \wedge \tilde{\mathcal{B}}^r $ on the boundary still does.
Let us consider now the path integral, if  we sum over  all integral classed associated to 
\begin{align}
    [\frac{d\mathcal{B}^r}{2\pi}]\in H^2(\Sigma_3,\mathbb{Z}).
\end{align}
(Recall that \eqref{eq:39} is only valid on-shell).
Then  the topological  contribution to  $$\exp \bigg(2\pi ik\int_{\mathcal{D}} \frac{d\mathcal{B}^r}{2\pi}\wedge \frac{\tilde{\mathcal{B}}^r}{2\pi}  \bigg),$$ arising from harmonic terms in the Hodge decomposition of $\mathcal{B}^r$ and $\tilde{\mathcal{B}}^r$ is of the form:
\begin{align}
    \exp (2\pi i kn m).
\end{align}
Hence, if we sum for all $n$ and use of the Poisson-summation formula, one gets:
\begin{align}
    \delta(km-l) \quad l\in \mathbb{Z},
\end{align}
with yields 
\begin{align}
    m=\frac{l}{k}, \quad k\neq 0.
\end{align}
That is, quantum mechanically  the flat $\mathcal{G}_1^{\nabla_c}$-gerbe, is projected onto its $k-$torsion sector, 
\begin{align}
    [\frac{\mathcal{\Tilde B}^r}{2\pi}]\in H^2(\Sigma_3,\mathbb{Z}_k) \quad \textrm{on-shell}.
\end{align}
Besides, since $dX^r=k\mathcal{B}^r$ and $[dX^r/2\pi k]\in H^1(\Sigma_3,\mathbb{Z})$, then the class $[\mathcal{B}^r]$ belongs to
\begin{align}
    [\frac{\mathcal{B}^r}{2\pi}]\in H^1(\Sigma_3,\mathbb{Z}).
\end{align}
This proves the breaking of the $U(1)$ gauge symmetry into a $\mathbb{Z}_k$ subgroup. We now show how the holonomy transform under the $\mathbb{Z}_k$ transformation.

\paragraph{$\mathbb{Z}_k$ transformation acting on the holonomies.}
Let us define the holonomy associated with the gerbe with connection $\tilde{B}^r$
\begin{align}
 \text{Hol}_{\nabla}(\mathcal{N}):=\exp( i\oint_{\mathcal{N}}\mathcal{\tilde{B}}^r),
\end{align}
The action $S_T$ \eqref{ECUACION: ACTION FREE ANOMALY} is invariant under the transformation $\tilde{\mathcal{B}}^r+\chi^r$, where $\chi^r$ represents the connection of a k-torsion flat gerbe. The transformation of the holonomy under this change is 
\begin{align}
 \text{Hol}_{\nabla}(\mathcal{N})\longrightarrow \exp( i\oint_{\mathcal{N}}\chi^r) \,\text{Hol}_{\nabla}(\mathcal{N}),
\end{align}
with
\begin{align}
  \exp( i\oint_{\mathcal{N}}\chi^r)=\exp(2\pi i \frac{l}{k})\,\in \, \mathbb{Z}_k
\end{align}
where $l$ is an integer number. Consequently, the holonomy transforms under a $\mathbb{Z}_k$ symmetry, that is, the original $U(1)$ symmetry has been broken into a $\mathbb{Z}_k $ subgroup.

\section{Torsional sector of M2-brane}\label{CHAPTER: TORSIONAL SECTOR OF M2-BRANE}

In this section we consider a $\Sigma_3$ worldvolume of the bosonic M2-brane whose global description  is a (non-trivial) $T^2$-bundle over $S^1$, with torsion, where the $S^1$ dimension is associated to the euclidean time coordinate.
The manifold corresponds to a three-dimensional  \textit{twisted} torus with parabolic monodromy.
We consider an action that does not include a coupling to a WZ term containing the pullback of the Supergravity $C_{[3]}$, since we are only interested in the topological properties of the gauging procedure, without involving any Supergravity dynamics.

\noindent
Most of the results we obtained are also valid when one considers a flat $C_{[3]}$ connection, since in this case, the field equations  remain the same as we got in previous section.

\noindent
The monodromy of the twisted torus imposes an additional restriction on $\widetilde{\mathcal{B}}^r$ in order to be invariant under it. Besides this point the breaking to $\mathbb{Z}_k$ follows as in the section \ref{SECCION: M9T2 GERBE}

\noindent
Let $\pi:E\to S^{1}$ be a fiber bundle with fiber $T^{2}$, defined as the \emph{mapping torus} of a diffeomorphism
$\mathbb{A}\in SL(2,\mathbb Z)$:
\begin{align}
    E \cong \big(T^{2}\times [0,1]\big)\Big/\big((u,v,1)\sim (\mathbb{A}(u,v),0)\big).
\end{align}
In the parabolic case, up to conjugation we may take
\begin{align}\label{EQUATION: PARABOLIC MONODROMY}
    \mathbb{A}=\begin{pmatrix} 1 & k \\ 0 & 1 \end{pmatrix},
\qquad k\in\mathbb Z.
\end{align}
In coordinates $(u,v)\in \mathbb R^{2}/\mathbb Z^{2}$ on $T^{2}$, the monodromy acts as
\begin{equation}
\mathbb{A}(u,v)=(u+k\,v,\;v)\qquad (\mathrm{mod}\;1).
\end{equation}
In a torus bundle formulation of the M2-brane, the structure group is the area preserving diffeomorphisms which is isomorphic to the symplectomorphisms (since we are in two-dimensions). The monodromy is defined as a map 
$$\pi_1(\Sigma_2)\to \pi_0(Symp(T^2))=SL(2, \mathbb{Z})$$
That is a map from the fundamental group of the base manifold to the group of isotopy classes of the structure group of the fiber. See for example, \cite{Kahn2004SymplecticTB}, \cite{GarciadelMoral:2011av}.}

\noindent
Equivalently, we may write the total space as
\begin{align}
    E \cong{}& \big(T^{2}\times [0,1]\big)\Big/\big((u,v,1)\sim (u+k\,v,\;v,\;0)\big).
\end{align}
The homotopy classes of global sections $s:S^{1}\to E$ are classified by cohomology with twisted coefficients
\begin{equation}
H^{1}\!\left(S^{1},\mathbb{Z} \oplus \mathbb{Z}_{\mathbb{A}}\right),
\end{equation}
where $\mathbb Z\oplus \mathbb{Z}_{\mathbb{A}}$ denotes the local system (a $\mathbb Z$-module) given by $\mathbb{Z} \oplus \mathbb{Z}$ with the
$\pi_{1}(S^{1})\cong \mathbb Z$ action induced by $\mathbb{A}$.
In particular \cite{delMoral:2025xhj}, for $S^{1}$ one has the standard isomorphism
\begin{equation}
H^{1}\!\left(S^{1},\mathbb{Z} \oplus \mathbb{Z}_{\mathbb{A}}\right)\;\cong\;
\mathbb{Z} \oplus \mathbb{Z}/(\mathbb{A}-id_{SL(2,\mathbb{Z})})\mathbb{Z}\oplus\mathbb{Z}.
\end{equation}
For the parabolic monodromy $\mathbb{A}$,
\begin{align}
\mathbb{A}-id_{SL(2,\mathbb{Z})}=\begin{pmatrix}0&k\\0&0\end{pmatrix}, \quad
(\mathbb{A}-id_{SL(2,\mathbb{Z})})\binom{q}{p'}=\binom{kp'}{0}.
\end{align}
Hence
\begin{align}
(\mathbb{A}-id_{SL(2,\mathbb{Z})})~\mathbb{Z}\oplus\mathbb{Z}
=\{(kq,0):q\in\mathbb Z\}
=k\mathbb Z\oplus\{0\},
\end{align}
and therefore
\begin{align}
    H^{1}\!\left(S^{1},\mathbb{Z} \oplus \mathbb{Z}_{\mathbb{A}}\right)
\;\cong\;
\mathbb{Z} \oplus \mathbb{Z}/(k\mathbb Z\oplus\{0\})
\;\cong\;{}&
\mathbb Z\oplus\mathbb Z_{k},
\\
{}&\quad (k\neq 0)\notag.
\end{align}
The $\mathbb Z$ factor records the \emph{winding number} along the invariant direction in the fiber (since $\mathbb{A}$ fixes the
$v$ coordinate), while the $\mathbb Z_{k}$ factor measures the discrete obstruction associated with the \emph{shear}.
In the special cases:
\begin{itemize}
\item If $k=1$, then $\mathbb Z\oplus \mathbb Z_{1}\cong \mathbb Z$ (there is no nontrivial discrete label).
\item If $k=0$, the monodromy is trivial  and via the \textit{universal coefficient theorem:}
\begin{equation}
H^{1}(S^{1},\mathbb Z\oplus\mathbb{Z})\cong \mathbb{Z}\oplus\mathbb{Z},
\end{equation}
i.e. sections are classified by two winding numbers.
\end{itemize}
The homotopy classes of global sections are classified in $k$ \emph{families} labeled by $\mathbb Z_{k}$, within each family there are infinitely many classes indexed by $\mathbb Z$. In particular, \emph{each torsion family} contains a constant section.
Indeed, a constant section $s(t)=(\hat u,\hat v)\in T^{2}$ must satisfy the gluing condition
\begin{equation}
A(\hat u,\hat v)=(\hat u,\hat v)\quad \text{in }T^{2},
\end{equation}
which is equivalent to
\begin{equation}
\hat u+k\hat v\cong \hat u ~~ (\mathrm{mod}\;1)
\qquad\Longleftrightarrow\qquad
k\hat v\in\mathbb Z.
\end{equation}
Therefore
\begin{align}
    \hat v=\frac{r}{k}\ (\mathrm{mod}\;1),
\qquad r\in\mathbb Z_{k},
\qquad \hat u\ \text{arbitrary},
\end{align}
where $r$ labels a constant section within each torsion family. 
\newline
As shown in \cite{Brylinski:1993ab}, a $\mathcal{G}_1^\nabla$-gerbe with connection is described, on a good open cover $\{U_i\}_{i\in I}$, by Čech--Deligne local data
\begin{equation}
\bigl(\widetilde{\mathcal{B}}_i,\;A_{ij},\;g_{ijk}\bigr),
\end{equation}
where $\widetilde{\mathcal{B}}_{i}\in\Omega^{2}(U_{i})$, $A_{ij}\in\Omega^{1}(U_i\cap U_j)$, and $g_{ijk}:U_i\cap U_j\cap U_k\to U(1)$ satisfy the usual gerbe cocycle relations. The global curvature is a closed 3-form $F_{[3]}$, locally given by $F_{[3]}=d\widetilde{\mathcal{B}}_{i}$ in $U_i$. 
The classes of flat gerbes are classified by
\begin{align}
    H^{2}(E;U(1))\;\cong\;H^{2}(E;\mathbb R/\mathbb Z),
\end{align}
and there is a canonical short exact sequence
\begin{align}
0\;\rightarrow\;\frac{H^{2}(E;\mathbb R)}{H^{2}(E;\mathbb Z)}
\;\rightarrow\;
H^{2}(E;\mathbb R/\mathbb Z)
\;\rightarrow\;
\mathrm{Tor}\,H^{3}(E;\mathbb Z)
\;\rightarrow\;0,
\end{align}
so that the torsion subgroup of $H^{3}(E;\mathbb{Z})$ controls the discrete part of flat classes \cite{Brylinski:1993ab}. Consequently, a $\mathcal{G}_1^{\nabla_c}$-gerbe is completely determined by its connection, which allows us, in what follows, to analyze its Holonomy and to make explicit how the torsion class acts in the \textit{winding} sector.
Let $t$ be a coordinate on the base circle, $t\sim t+1$, so that
$t\mapsto e^{2\pi i t}\in S^{1}$ and $dt$ is globally defined.
On the fiber $T^{2}$ we use coordinates $(u,v)$ with identifications
\begin{equation}
(u+1,v)\sim(u,v),\qquad (u,v+1)\sim(u,v).
\end{equation}
The monodromy \eqref{EQUATION: PARABOLIC MONODROMY} induces the identification on 1-forms
\begin{equation}
(du,dv)\sim (du+k\,dv,\;dv).
\end{equation}
We therefore seek a 2-form connection $\widetilde{\mathcal{B}}$ invariant under this identification. A natural choice is
\begin{equation}
\widetilde{\mathcal{B}}=\frac{m}{k}\,dt\wedge dv,
\qquad m=1,\dots,k-1,
\end{equation}
which is globally defined on $E$ and satisfies
\begin{equation}
d\widetilde{\mathcal{B}}=0,
\end{equation}
hence it defines the $\mathcal{G}_1^{\nabla_c}$-gerbe. The topological information of the $\mathcal{G}_1^{\nabla_c}$-gerbe can be captured by its holonomies (Wilson surfaces). For a closed surface
$\Sigma \subset E$,
\begin{align}
\mathrm{Hol}_\nabla(\Sigma)=\exp\!\left(2\pi i\int_{\Sigma}\widetilde{\mathcal{B}}\right),
\end{align}
and for a 2-surface sweeping the $(t,v)$ directions with
$\langle dt, dv\rangle =1$ \footnote{Where we used that $\langle a, b\rangle=\int_\Sigma a \wedge b$.}, we obtain
\begin{align}
    \mathrm{Hol}_\nabla(\Sigma)=\exp \left(2\pi i\,\frac{m}{k}\right),
\end{align}
which is a $k$-th root of unity (identifiable with an element of $\mathbb Z_{k}$). 
\newline
Now, let $S^{1}_{v}$ be the circle in the $v$ direction with $\int_{S^{1}_{v}} dv=1$. The 2-form $\widetilde{\mathcal{B}}$ determines a
 $\Xi$ 1-form on the base by fiber integration (transgression map)
\begin{align}\label{eq:73}
    \Xi={}&\int_{S^{1}_{v}}\widetilde{\mathcal{B}}\\
={}&\int_{S^{1}_{v}}\frac{m}{k}\,dt\wedge dv\\
={}&\frac{m}{k}\,dt.
\end{align}
This $\Xi$ defines a flat principal $U(1)$-bundle connection on $S^{1}$, whose topological invariant is the Holonomy
\begin{align}
\exp \left(i\oint_{S^{1}}\Xi\right)={}&
\exp\!\left(i\int_{0}^{1}\frac{m}{k}\,dt\right)\\
={}&\exp \left(2\pi i\,\frac{m}{k}\right)~ \in ~\mathbb Z_{k},
\end{align}
where Tor$H^3(\Sigma_3,\mathbb{Z})\cong \mathbb{Z}_k$, making manifest the role of the torsion in the generating non trivial Holonomy of the $\mathcal{G}_1^{\nabla_c}$-gerbe, \textit{i.e.} Wilson surfaces, Wilson loops.

\subsection{The coupling of the winding current to the gerbe connection.}\label{CHAMPTER: The coupling of the winding current to the gerbe connection}
We consider an open cover $\{V_0,V_1.V_2\}$ of $E$ coming form three intervals of $S^1$, $\{U_0,U_1,U_2\}$, where for simplicity, there are not triple overlaps. We define the Deligne data, 
\begin{align}
\widetilde{\mathcal{B}}_0=\widetilde{\mathcal{B}},\quad \widetilde{\mathcal{B}}_1=\widetilde{\mathcal{B}}+\frac{m}{k}f^{'}dt\wedge dv, \quad \widetilde{\mathcal{B}}_2=\widetilde{\mathcal{B}},\quad  d\widetilde{\mathcal{B}}=0,
\end{align}
where $f(t) \in \mathcal{C}^{1}(S^1)$ is a function
\begin{align}
     f:S^1\to\mathbb{R}
 \end{align}
such that $f^{'}(t)=df/dt$ is not identically zero and $supp(f^{'})\in U_0\cap U_1$.
 $f(t)$ is a continuous differentiable function on the universal cover $\mathbb{R}\to S^1=\mathbb{R}/\mathbb{Z}$ with a periodic $f(t)^{'}$. $f(t)^{'}$ is a well-defined function over $S^1$, while $f(t)$ is a multivalued function over $S^1.$ Since there are not triple overlaps, the \textit{Čech--Deligne} equations are:
\begin{align}
     &\widetilde{\mathcal{B}}_1-\widetilde{\mathcal{B}}_0=\frac{m}{k}f^{'}dt\wedge dv=dA_{01}\quad in \quad V_0\cap V_1, \\
     &\widetilde{\mathcal{B}}_2-\widetilde{\mathcal{B}}_1=0 \quad in \quad V_1\cap V_2, \\
      &\widetilde{\mathcal{B}}_0-\widetilde{\mathcal{B}}_2=0 \quad in \quad V_2\cap V_0.
 \end{align}
Where $A_{01}$ is defined as, $A_{01}=\frac{m}{k}f dv$. Although $f$ is a multivalued function over $S^1$, $dA_{01}$ is a well defined 1-form.  For the construction of $\check{C}$\textit{ech-Deligne} cohomology it is not necessary for $f$ to be defined globally $S^1$.
Also the curvature is zero $d\widetilde{\mathcal{B}}_i=0$, with $i=0,1,2$, we then have a flat gerbe connection.

\noindent
The unique topological invariant (in the sense of a cohomology class) associated with the connection is its Holonomy \cite{Johnson:2002tc}.
\begin{align}
 Hol_{\nabla}(\Sigma)={}&\exp \bigg(2\pi i\left[\sum_i\int_{\Sigma_i}\widetilde{\mathcal{B}}_i-\sum_{i<j}\int_{\Sigma_{ij}}A_{ij}\right]  \bigg)\\
 ={}&\exp(2\pi i \int_{\Sigma}\widetilde{\mathcal{B}})\\
 ={}&\exp(2\pi i \frac{m}{k})\in \mathbb{Z}_k,
 \end{align}
 where $\Sigma$ is a closed oriented surface, which decomposes into $\Sigma_i\subset V_i$ so that overlaps occur only in pairs $\Sigma_{ij}=\Sigma_i\cap\Sigma_j$.
\section{Algebra and charged operator for HSF of bosonic membrane}\label{CHAPTER: ALGEBRA}
The operator algebra implied by the path-integral Ward identity takes the form:
\begin{align}
    \langle U_{\alpha}^{}(\mathcal{N}) U_{\beta}^{}(\mathcal{Q}) \rangle \cong{}&  \exp\big(i\frac{ab}{k} I(\mathcal{N},\mathcal{Q}) \big)
    \langle U_{\beta}^{}(\mathcal{Q}) U_{\alpha}^{}(\mathcal{N}) \rangle.
\end{align}
Where $a,b\in \mathbb{Z}$. Here, $I(\mathcal{N},\mathcal{Q})$ denotes the intersection number between $\mathcal{N}$ and $\mathcal{Q}$, which is an integer different from zero $I(\mathcal{N},\mathcal{Q})\in\mathbb{Z}^{\times}$. Since there are no more topological operators in the theory, this is the only non-trivial relationship between the operators:
\begin{align}
\mathcal{U}_{\alpha}^r(\mathcal{N})={}&\exp \bigg(i\alpha\int_{\mathcal{N}} \star DX^r\bigg),
\\\mathcal{U}_{\beta}^r(\mathcal{Q})={}&\exp \bigg(i\beta\int_{\mathcal{Q}}  DX^r\bigg),
\end{align}
where $DX^r=dX^r-k\mathcal{B}^r$.
We now consider the \textit{Vortex} operator given by $\mathcal{O}^r_l(\{pt\})=\exp \big( il X^r\big)$ with $l \in \mathbb{Z}^\times$ acting on points $\{pt\}$. This operator is charged with respect to HFS$(0)$, generated by \(U_{\alpha}(\mathcal{N})\): when the the global symmetry is gauged, this operator is not  invariant. This allows to define a new operator that we denote as, \textit{Vortex-dressed} operator:
\begin{align}
\mathcal{O}_{l}(\{pt\},\gamma)={}&\exp(ilX^r)\exp(-ilk\int_{\gamma}\mathcal{B}_r)
\end{align}
where $\gamma$ is a open curve\footnote{We consider to $\epsilon^r(0)=0$}. The operator $\mathcal{O}_l$ is point-like\footnote{This operator can be naturally constructed within the framework of relative cohomology
$H^\bullet(\mathcal{M},\mathcal{N};\mathbb{Z})$ \cite{Becker:2013RelativeDifferential}. However, a detailed treatment of this construction lies beyond the scope of the present work.}, and the \textit{monopole} operator acts on it as follows:
\begin{align}
    \langle U_{\alpha}(\mathcal{N}) \mathcal{O}_l(\{ pt\},\gamma)\rangle \cong{}& 
    \exp \big( 2\pi il \alpha \text{Link}\langle \mathcal{N} | \{pt\}
    \rangle \big) \cdot \cdots\notag
    \\
    {}&\cdot \langle \mathcal{O}_l(\{ pt\},\gamma) \rangle,
\end{align}
where the linking number $\text{Link}\langle  \mathcal{N} | \{ pt\} \rangle \in \mathbb{Z}^\times$, and $\alpha=m/k$ with $m\in \mathbb{Z}.$ 

\noindent
Let us consider the information captured by the $\mathcal{G}_1^{\nabla_c}$-gerbe, which is globally encoded in  Hol$_{\nabla} (\mathcal{C}_2)$ \cite{Brylinski:1993ab} and \cite{Baez:2010ya}, this topological invariant which is a Wilson Surface, becomes nontrivial due to the action of the background fields acting on the M2-brane,
\begin{align}
    \mathcal{W}(\mathcal{C}_{(2)}) = \exp \big( iq \int_{\mathcal{C}_{2}} \widetilde{\mathcal{B}}\big),
\end{align}
where \(\mathcal{C}_2\) is a 2-cycle basis of  a submanifold of codimension one and with a quantized charge \(q \in \mathbb{Z}^\times\) \cite{Brennan:2022tyl}.\\

\noindent
The Wilson surface on the other hand is a natural object to be charged with respect to the gauge symmetry given by  \eqref{SYMMETRIA GLOBAL MEMBRANA ANAISADA}.

\noindent
Using the transgression of the $\mathcal{G}_1^\nabla$-gerbe with two-form potential $\widetilde{\mathcal{B}}$ along the fiber cycle, see for example \eqref{eq:73}. Consequently, the holonomy along the one-cycle $\Gamma$ must be understood as the holonomy under transgression induced $U(1)$ connection, rather than as the direct holonomy of the original gerbe. We denoted this by $Hol_{\nabla}^{tr}(\Gamma)$ 
\begin{align}
   \langle U_{\beta}(\mathcal{Q}) Hol_{\nabla}^{tr}(\Gamma) \rangle \cong{}& \exp \big( i \beta \text{Link}\langle \mathcal{Q} ~|~ \Gamma
    \rangle \big) \langle Hol_{\nabla}^{tr}(\Gamma)  \rangle,
\end{align}
with $\Gamma$ a one-cycle and $\beta=n/k$ with $n\in \mathbb{Z}$. This, however, does not exclude the existence of another formalism in which   the charged object is directly the Wilson Surface in terms of a generalized version of the Linking number including  the torsional effects like torsion linking forms \cite{Hillman:2011Linking}. The existence of torsion structures has been shown in detailed the preceding section. Also in \cite{delMoral:2025xhj}, different torsional structures on the  supersymmetric M2-brane with fluxes were identified. This point is out of the scope of the present article and will be consider elsewhere.

\section{Conclusions}\label{CHAPTER: CONCLUSIONS}
In this paper we study the interplay between higher-form symmetries, anomaly cancellation, gerbe structures and flux quantization conditions for the bosonic M2-brane. 

\noindent
We have shown that a consistent implementation of higher form symmetries (HFS) for the Polyakov action, once the global symmetry is gauged, requires the introduction of higher-degree background fields coupled through a BF-type term. The theory has a mixed t' Hooft anomaly; its cancellation requires the addition of an inflow term on a four-manifold bounding the worldvolume. Within this framework, we constructed the topological operators associated with the HFS and computed their operator algebra. In particular, the global structure of the M2-brane theory given by gerbe structures are naturally characterized by Wilson surface operators capturing the global holonomy of the membrane.
\newline
A relevant consequence is that the symmetry breaking can be understood as a direct result of the  presence of a $U(1)$ gerbe-bundle structure over $\Sigma_3$. On-shell, the  U(1) Gerbe-bundle ($\mathcal{G}_1^{\nabla}$-gerbe) is projected into a torsion class $\mathbb{Z}_k$. 
The existence of the $\mathcal{G}_1^{\nabla}$-gerbe induces a non-trivial worldvolume flux condition. It has been proved that such a flux condition, which may arise from different mechanisms, implies discreteness of the mass operator spectrum in the supersymmetric theory on toroidal backgrounds~\cite{Boulton:2002br}. The supersymmetric formulation of the M2-brane is Lorentz anomaly free in $D=11$ and is therefore consistent from this viewpoint. In the supersymmetric theory, a worldvolume flux condition may appear either from an irreducible wrapping that induces a central-charge condition~\cite{Martin:1997cb}, or from the presence of a flat quantized 3-form, (i.e. it has a vanishing curvature, $G_{[4]} =dC_{[3]} =0$. Its pullback induces an $F_{[2]} $ flux on the M2-brane worldvolume~\cite{GarciaDelMoral:2018jye}, \cite{GarciadelMoral:2020dfs}. Indeed, for certain flux components one can identify a duality between these two realizations~\cite{GarciaDelMoral:2018jye}. In this paper, the background fields are dynamical fields. One could conjecture a bulk origin of the background fields, associated to the pullback of the supergravity flat connection $C_{[3]} $ of a  $\mathcal{G}_{2}^{\nabla_c}$-gerbe, projected on the $T^2$ of the target space leading to $\mathcal{B}^r$  $U(1)$ connections and when projected on the $S^1\subset T^2$ leading to  $\widetilde{\mathcal{B}}^r$ 2-connections, that couple to the embedding maps.  The contribution of the pullback of this particular 3-form coupling to the action, does not alter the equations of motion, not the computations done in the paper. It only requires compatibility with the topological structure of the background fields. The term contributes as a topological term to the action and is consistent with a Ricci flat metric like the one associated with the target here considered, it satisfies the 11d Supergravity equations of motion hence it is a consistent physical interpretation. In any case we will consider this possibility in more detail elsewhere.
\newline
Finally, the most general bosonic action includes the Wess--Zumino term
\begin{equation*}
S_{M2}(g,X^M)+S_{\mathrm{WZ}}(X^M,C_{[3]})~~\supset~~\frac{1}{2}\int_{\Sigma_3}\mathbb{P}(C_{[3]}),
\end{equation*}
where $\mathbb{P}(C_{[3]})$ denotes the pullback of the Supergravity three-form.
This theory, analogous to a \textit{Chern-Simons} type coupling, exhibits an ABJ-type anomaly on the worldvolume~\cite{Montero:2017yja}. This anomaly can be studied by constructing defect operators $\mathcal{D}_{\alpha}(\mathcal{N})$, which in turn calls for introducing non-trivial TQFT-type actions supported on the defect in $\Sigma_3$ (see~\cite{Garcia-Valdecasas:2023mis}, \cite{Choi:2022jqy},\cite{Karasik:2022kkq}, \cite{Hasan:2024aow}, \cite{Fernandez-Melgarejo:2024ffg} and \cite{Cvetic:2023plv}). The implications of this mechanism for the quantum formulation of the M2-brane are not developed here and are left for future work.

\section*{Acknowledgements}
We thank to C. Las Heras for helpful discussion. FCP is grateful to the physics Departament of Pontificia Universidad Católica de Chile,  Chile. FCP is supported by Doctorado nacional (ANID) 2023 Scholarship N$21230379$ and supported as graduate student in the “Doctorado en Física Mención Física-Matemática” Ph.D. program at the Universidad de Antofagasta. MPGM is grateful to the Physics Department of Sciences Faculty at the University of Antofagasta, Chile, for their kind invitation, where part of this work was done.  MPGM has been  partially supported by the PID2024-155685NB-C21 MCI Spanish Grants and by the University of La Rioja project REGI2025/41. AR and  want to thank to SEM18-02 project of the U. Antofagasta and protect MATH-AMSUD 24-MAT-12D and the Scientific Research Computing Institute of the University of La Rioja (SCRIUR), Spain.

\begin{appendix}
\numberwithin{equation}{section}

\section{Invariance of action with background fields}\label{APENDICE DEMOSTRACION FREE ANOMLA TERM} 
 The kinetic term in the action \eqref{ECAUCION: ACCION LIBRE DE ANOMALIA} is invariant, while for the second term we have, 
\begin{align}
 \delta \int_{\Sigma_3}\Tilde{\mathcal{B}}^r\wedge dX^r={}&\int_{\Sigma_3}\Tilde{\mathcal{B}}^r\wedge d\epsilon^r.
\end{align}
Now, with the inflow term:
\begin{align}
    \delta \big( \frac{k}{2} \int_{\Sigma_3}\Tilde{\mathcal{B}}^r\wedge dX^r+\mathcal{T}_{TQFT}\big)={}& \frac{k}{2}\int_{\Sigma_3}\Tilde{\mathcal{B}}^r\wedge d\epsilon^r-\notag
    \\
    {}&-\frac{k}{2}\int_{\mathcal{D}}d( d\epsilon^r \wedge \Tilde{\mathcal{B}}^r)\\  
    ={}&0. \notag
\end{align}
Via Stokes' theorem and \(\partial \mathcal{D} = \Sigma_3\), the second integral is defined on \(\Sigma_3\) and cancels the anomaly term. 
\subsection{Cancellation of the anomaly in the compact sector}
Our proposal for the anomaly action with \( \text{supp}(\mathcal{D}) \) is given by:
\begin{align}
    \mathcal{T}_{TQFT} = \frac{k}{2} \int_{\mathcal{D}} \mathcal{B}^r \wedge d\tilde{\mathcal{B}}^r.
\end{align}
For the non-trivial class of the compact sector given by \( [\mathcal{B}^r] \) and \( [\tilde{\mathcal{B}}^r] \), we have the following:
\begin{align}
    \mathcal{T}_{TQFT}' = \frac{k}{2}\int_{\mathcal{D}} d\epsilon^r \wedge d\tilde{\mathcal{B}}^r + \mathcal{T}_{TQFT}.
\end{align}
Thus, in the path integral \(\tilde{\mathcal{Z}}\), transforming under the equivalence class of the background field gives:
\begin{align}
    \delta_{[\Upsilon]}\tilde{\mathcal{Z}} ={}& \mathcal{Z}\exp(\mathcal{T}_{TQFT}') \notag \\ 
    = {}& \mathcal{Z} \, \exp\big( \frac{k}{2}\int_{\Sigma_3} d\epsilon^r \wedge \tilde{\mathcal{B}}^r\big)   \exp(\mathcal{T}'_{TQFT}) \notag \\ \notag
    = {}& \mathcal{Z} \,\exp\big( \frac{k}{2}\int_{\Sigma_3} d\epsilon^r \wedge \tilde{\mathcal{B}}^r\big) \exp(\mathcal{T}_{TQFT})  \cdot \notag
    \\
    {}& \cdot\exp  \big(\frac{k}{2}\int_{\mathcal{D}} d\epsilon^r \wedge d\tilde{\mathcal{B}}^r \big) \notag \\
    = {}& \mathcal{Z}\,  \exp(\mathcal{T}_{TQFT}) \notag \\
    = {}& \tilde{\mathcal{Z}}.
\end{align}
Where $[\Upsilon]:= ([X^r], [\mathcal{B}^r], [\tilde{\mathcal{B}}^r])$.
In the fourth line, we perform integration by parts on $\delta \mathcal{T}_{TQFT} $, which results in the cancellation of the anomaly term in \( \Sigma_3 \). With this argument, we demonstrate that the compact sector of the theory is free of anomalies, provided that a \( TQFT \) term is introduced, supported on a 4-manifold whose boundary coincides with the worldvolume \( \Sigma_3 \) \cite{Garcia-Valdecasas:2023mis}.
\section{Independence of the theory on the auxiliary manifold $\mathcal{D}$ }\label{APENDICE: INDEPENDENCIA DEL BORDISMO}
\renewcommand{\theequation}{B.\arabic{equation}}
We prove in this section that the four dimensional manifold $\mathcal{D}$ such that $\partial\mathcal{D}=\Sigma_3$ is non-physical, since the topological term do not depend on the choice of the auxiliary manifold. 
Assume that there exists two different four auxiliary manifolds $\mathcal{D}$ and $\mathcal{D}^{'}$ with the same boundary $\Sigma_3$. It is possible to form a closed 4-manifold 
\begin{align}
\mathcal{Y}_4=\mathcal{D}\cup_{\Sigma_3} (-\mathcal{D}')
\end{align}
The independence of the theory from $\mathcal{D}$ appears when the path integral contribution in the original and deformed manifold are compared, 
\begin{align}
\exp({-{S_{bulk}(\mathcal{D}}}))\cong{}&\exp\bigg(-\frac{i k}{2\pi}\int_{\mathcal{Y}_4} \mathcal{B}\wedge d\tilde{\mathcal{B}}\bigg) ~ \exp\big({S_{bulk}[\mathcal{D}^{'}]}\big)
\end{align}
and 
\begin{align}\label{indep} exp\bigg(-\frac{i k}{2\pi}\int_{\mathcal{Y}_4} \mathcal{B}\wedge d\tilde{\mathcal{B}} \bigg)=1
\end{align} 
This condition is satisfied by imposing to the background fields to be in the integer cohomology on $\mathcal{Y}_4$
\begin{align}
\bigg[\frac{1}{2\pi}\mathcal{B}\bigg] \in H^1(\mathcal{Y}_4,\mathbb{Z}), \quad \bigg[\frac{1}{2\pi}d\tilde{\mathcal{B}}\bigg] \in H^3(\mathcal{Y}_4,\mathbb{Z}),
\end{align}
hence with this conditions it is automatically guaranteed that $\int_{\mathcal{Y}_4}\mathcal{B} \wedge d\tilde{\mathcal{B}}\in \mathbb{Z}$, satisfying the condition \eqref{indep}. These are the conditions that the background fields of our theory satisfy. In section \ref{SECCION: M9T2 GERBE} when we take a k-level BF-theory since $\Tilde{\mathcal{B}}^r$ is a connection with a $k$ torsion class, $[\frac{k}{2\pi}d\Tilde{\mathcal{B}}]\in H^3(\mathcal{Y}_4,\mathbb{Z}).$
\end{appendix}

%
%

\end{document}